\documentclass[final,12pt]{article}
\usepackage{hyperref}
\usepackage{amssymb,amsmath,amsthm}
\usepackage{enumerate}
\usepackage[conditional,light,first,bottomafter]{draftcopy}
\draftcopyName{DRAFT\space\today}{130}
\draftcopySetScale{65}
\usepackage[letterpaper,hmargin=3.3cm,vmargin={2.6cm,3.2cm}]{geometry}
\usepackage{setspace}
\makeatletter
\renewcommand{\section}{\@startsection%
{section}%
{1}%
{0em}%
{1.7em}%
{1.2em}%
{\normalfont\large\centering\bfseries}}
\renewcommand{\@seccntformat}[1]%
{\csname the#1\endcsname.\hspace{0.5em}}
\makeatother

\numberwithin{equation}{section}

\newtheorem{theorem}{Theorem}
\newtheorem{proposition}{Proposition}
\newtheorem{lemma}{Lemma}

\theoremstyle{definition}

\newtheorem{remark}{Remark}

\newtheorem*{acknowledgments}{Acknowledgments}

\newcommand{\abs}[1]{\left|#1\right|}
\newcommand{\norm}[1]{\left\|#1\right\|}
\newcommand{\I}{{\rm i}}

\def\cprime{$'$}
\def\ocirc#1{\ifmmode\setbox0=\hbox{$#1$}\dimen0=\ht0 \advance\dimen0
  by1pt\rlap{\hbox to\wd0{\hss\raise\dimen0
  \hbox{\hskip.2em$\scriptscriptstyle\circ$}\hss}}#1\else {\accent"17 #1}\fi}

\DeclareMathOperator{\dom}{Dom}
\DeclareMathOperator{\Ker}{Ker}

\DeclareMathOperator{\Sp}{\sigma}

\begin{document}

\title{The Two-Spectra Inverse Problem for Semi-Infinite\\
       Jacobi Matrices in The Limit-Circle Case%
\footnotetext{%
Mathematics Subject Classification(2000):
47B36, % Jacobi operators and generalizations
49N45, % Inverse problems
81Q10, % Self-adjoint operator theory in quantum theory, including
       % spectral analysis.
47A75, % Eigenvalue problems
47B37, % Operators on special spaces (weighted shifts, operators on
       % sequence spaces, etc)
47B39}%% Difference operators
\footnotetext{%
Keywords:
Jacobi matrices;
Two-spectra inverse problem;
Limit circle case}%
\footnotetext{%
Research partially supported by CONACYT under Project P42553­F.}
\\[6mm]}
\author{\textbf{Luis O. Silva%
\thanks{%
Author partially supported by PAPIIT-UNAM through grant IN-111906.}\,\,
and Ricardo Weder%
\thanks{%
Fellow Sistema Nacional de Investigadores.}}
\\[6mm]
%% ----- Institution ________
\small Departamento de M\'{e}todos Matem\'{a}ticos y Num\'{e}ricos\\[-1.6mm]
\small Instituto de Investigaciones en Matem\'aticas Aplicadas y en Sistemas\\[-1.6mm]
\small Universidad Nacional Aut\'onoma de M\'exico\\[-1.6mm]
\small C.P. 04510, M\'exico D.F.\\[1mm]
\small \texttt{silva@leibniz.iimas.unam.mx}\\[-1mm]
\small \texttt{weder@servidor.unam.mx}}
%%%%%%%%
\date{}
\maketitle
\begin{center}
\begin{minipage}{5in}
  \centerline{{\bf Abstract}} \bigskip We present a technique for
  reconstructing a semi-infinite Jacobi operator in the limit circle
  case from the spectra of two different self-adjoint
  extensions. Moreover, we give necessary and sufficient conditions for two real
  sequences to be the spectra of two different self-adjoint extensions
  of a Jacobi operator in the limit circle case.
\end{minipage}
\end{center}
\newpage

\section{Introduction}
\label{sec:intro}
In the Hilbert space $l_2(\mathbb{N})$, consider the operator $J$
whose matrix representation with respect to the canonical basis
in $l_2(\mathbb{N})$ is the semi-infinite Jacobi matrix
\begin{equation}
  \label{eq:jm-0}
  \begin{pmatrix}
    q_1 & b_1 & 0  &  0  &  \cdots
\\[1mm] b_1 & q_2 & b_2 & 0 & \cdots \\[1mm]  0  &  b_2  & q_3  &
b_3 &  \\
0 & 0 & b_3 & q_4 & \ddots\\ \vdots & \vdots &  & \ddots
& \ddots
  \end{pmatrix}\,,
\end{equation}
where $b_n>0$ and $q_n\in\mathbb{R}$ for $n\in\mathbb{N}$.  This
operator is densely defined in $l_2(\mathbb{N})$ and $J\subset J^*$
(see Section~\ref{sec:preliminaries} for details on how $J$ is
defined).

It is well known that $J$ can have either $(1,1)$ or $(0,0)$ as its
deficiency indices \cite[Sec.\,1.2 Chap.\,4]{MR0184042},
\cite[Cor.\,2.9]{MR1627806}. By our definition (see
Section~\ref{sec:preliminaries}), $J$ is closed, so the case $(0,0)$
corresponds to $J=J^*$, while $(1,1)$ implies that $J$ is a
non-trivial restriction of $J^*$. The latter operator is always
defined on the maximal domain in which the action of the matrix
(\ref{eq:jm-0}) makes sense \cite[Sec.\,47]{MR1255973}.

Throughout this work we assume that $J$ has deficiency indices
$(1,1)$. Jacobi operators of this kind are referred as being in the
limit circle case and the moment problem associated with the
corresponding Jacobi matrix is said to be indeterminate
\cite{MR0184042,MR1627806}. In the limit circle case, all self-adjoint
extensions of a Jacobi operator have discrete spectrum
\cite[Thm.\,4.11]{MR1627806}. The set of all self-adjoint extensions
of a Jacobi operator can be characterized as a one parameter family of
operators (see Section~\ref{sec:preliminaries}).

The main results of the present work are Theorem~\ref{thm:recovering}
in Section~\ref{sec:recovering} and
Theorem~\ref{thm:necessary-sufficient} in
Section~\ref{sec:n-s-conditions}. In Theorem~\ref{thm:recovering} we
show that a Jacobi matrix can be recovered uniquely from the spectra
of two different self-adjoint extensions of the Jacobi operator $J$
corresponding to that matrix. Moreover, these spectra also determine
the parameters that define the self-adjoint extensions of $J$ for
which they are the spectra.  The proof of Theorem~\ref{thm:recovering}
is constructive and it gives a method for the unique
reconstruction. The uniqueness of this reconstruction in a more
restricted setting has been announced in \cite{MR1045318} without
proof. In Theorem~\ref{thm:necessary-sufficient} we give necessary and
sufficient conditions for two sequences to be the spectra of two
self-adjoint extensions of a Jacobi operator in the limit circle
case. This is a complete characterization of the spectral data for the
two-spectra inverse problem of a Jacobi operator in the limit circle
case.

In two spectra inverse problems, one may reconstruct a certain
self-adjoint operator from the spectra of two different rank-one
self-adjoint perturbations of the operator to be reconstructed. This
is the case of recovering the potential of a Schr\"{o}dinger
differential expression in $L_2(0,\infty)$, being regular at the
origin and limit point at $\infty$, from the spectra of two operators
defined by the differential expression with two different self-adjoint
boundary conditions at the origin
\cite{AW1,AW2,borg,MR1329533,MR1754515,MR0039895,MR0062904,MR0162996,%MR0036916,
  MR0058064}. Necessary and sufficient conditions for this inverse
problem are found in \cite{MR0162996}. Characterization of spectral
data of a related inverse problem was obtained in \cite{MR2355343}.
The inverse problem consisting in recovering a Jacobi matrix from the
spectra of two rank-one self-adjoint perturbations, was studied in
\cite{MR2164835,MR49:9676,MR499269,MR0221315,MR1643529,MR2091673}. A
complete characterization of the spectral data for this two-spectra
inverse problem is given in \cite{weder-silva}.

In the formulation of the inverse problem studied in the present work,
the aim is to recover a symmetric non-self-adjoint operator from the
spectra of its self-adjoint extensions, as well as the parameters that
characterize the self-adjoint extensions. There are results for this
setting of the two spectra inverse problem, for instance in
\cite{MR997788,MR0052004} for Sturm-Liouville operators, and in
\cite{MR1045318} for Jacobi matrices.

It is well known that self-adjoint extensions of symmetric operators
with deficiency indices $(1,1)$ can be treated within the rank-one
perturbation theory (cf. \cite[Sec. 1.1--1.3]{MR1752110} and, in
particular, \cite[Thm.\,1.3.3]{MR1752110}). Thus, both settings may be
regarded as particular cases of a general two-spectra inverse
problem. A consideration similar to this is behind the treatment of
inverse problems in \cite{MR0190761}.  For Jacobi operators, however,
the type of rank-one perturbations in the referred formulations of the
inverse spectral problem are different \cite{MR1752110}. Indeed, in
the setting studied in \cite{weder-silva}, one has the so-called
bounded rank-one perturbations \cite[Sec. 1.1]{MR1752110}. This means
that all the family of rank-one perturbations share the same domain. In
contrast the present work deals with singular rank-one perturbations
\cite[Sec. 1.3]{MR1752110}, meaning that every element of the family
of rank-one perturbations has different domain. Note that for
differential operators both settings involve a family of singular
rank-one perturbations.

The paper is organized as follows. In Section~\ref{sec:preliminaries},
we introduce Jacobi operators, in particular the class whose
corresponding Jacobi matrix is in the limit circle case. Here we also
present some preliminary results and lay down some notation used
throughout the text. Section \ref{sec:recovering} contains the
uniqueness result on the determination of a Jacobi matrix by the
spectra of two self-adjoint extensions. The proof of this assertion
yields a reconstruction algorithm. Finally in
Section~\ref{sec:n-s-conditions}, we give a complete characterization
of the spectral data for the two spectra inverse problem studied here.

\section{Preliminaries}
\label{sec:preliminaries}
Let $l_{fin}(\mathbb{N})$ be the linear space of sequences with a
finite number of non-zero elements. In the Hilbert space
$l_2(\mathbb{N})$, consider the operator $J$ defined for every
$f=\{f_k\}_{k=1}^\infty$ in $l_{fin}(\mathbb{N})$ by means of the
recurrence relation
\begin{align}
  \label{eq:recurrence-coordinates}
  (Jf)_k &:= b_{k-1}f_{k-1} + q_k f_k + b_k
f_{k+1}\,,\quad k \in \mathbb{N} \setminus \{1\}\,,\\
  \label{eq:initial-coordinates}
 (Jf)_1 &:=  q_1 f_1 + b_1 f_2\,,
\end{align}
where, for $n\in\mathbb{N}$, $b_n$ is positive and $q_n$ is
real. Clearly, $J$ is symmetric since it is densely defined and
Hermitian due to (\ref{eq:recurrence-coordinates}) and
(\ref{eq:initial-coordinates}). Thus $J$ is closable and henceforth we
shall consider the closure of $J$ and denote it by the same letter.

We have defined the operator $J$ so that the semi-infinite Jacobi
matrix (\ref{eq:jm-0}) is its matrix representation with respect to
the canonical basis $\{e_n\}_{n=1}^\infty$ in $l_2(\mathbb{N})$ (see
\cite[Sec. 47]{MR1255973} for the definition of the matrix
representation of an unbounded symmetric operator). Indeed, $J$ is the
minimal closed symmetric operator satisfying
\begin{equation*}
\begin{array}{l}
(Je_n,e_n)=q_n\,,\quad (Je_n,e_{n+1})=(Je_{n+1},e_n)
  =b_n\,,\\
(Je_n,e_{n+k})=(Je_{n+k},e_{n})=0\,,
\end{array}
\quad n\in\mathbb{N}\,,\,k\in\mathbb{N}\setminus\{1\}\,.
\end{equation*}
We shall refer to $J$ as the \emph{Jacobi operator} and to
(\ref{eq:jm-0}) as its associated matrix.

The spectral analysis of $J$ may be carried out by studying the
following second order difference system
\begin{equation}
  \label{eq:main-recurrence}
 b_{n-1}f_{n-1}+ q_nf_n+
  b_nf_{n+1}=
  \zeta f_n\,,\quad n>1\,,\quad \zeta\in\mathbb{C}\,,
\end{equation}
with the ``boundary condition''
\begin{equation}
  \label{eq:boundary}
  q_1f_1+b_1f_2=\zeta  f_1\,.
\end{equation}
If one sets $f_1=1$, then $f_2$ is completely determined by
(\ref{eq:boundary}). Having $f_1$ and $f_2$, equation
(\ref{eq:main-recurrence}) gives all the other elements of a sequence
$\{f_n\}_{n=1}^\infty$ that formally satisfies
(\ref{eq:main-recurrence}) and (\ref{eq:boundary}). Clearly, $f_n$ is
a polynomial of $\zeta$ of degree $n-1$, so we denote
$f_n=:P_{n-1}(\zeta)$. The polynomials $P_n(\zeta)$, $n=0,1,2,\dots$,
are referred to as the polynomials of the first kind associated with
the matrix (\ref{eq:jm-0}) \cite[Sec.\,2.1 Chap.\,1]{MR0184042}.

The sequence $P(\zeta):=\{P_{k-1}(\zeta)\}_{k=1}^\infty$ is not in
$l_{fin}(\mathbb{N})$, but it may happen that
\begin{equation}
 \label{eq:generalized-eigenvector}
  \sum_{k=0}^\infty\abs{P_k(\zeta)}^2<\infty\,,
\end{equation}
in which case $P(\zeta)\in\Ker (J^*-\zeta I)$.

The polynomials of the second kind
$Q(\zeta):=\{Q_{k-1}(\zeta)\}_{k=1}^\infty$ associated with the matrix
(\ref{eq:jm-0}) are defined as the solutions of
\begin{equation*}
    b_{n-1}f_{n-1} + q_n f_n + b_nf_{n+1} = \zeta f_n\,,
\quad n \in \mathbb{N} \setminus \{1\}\,,
\end{equation*}
under the assumption that $f_1=0$ and $f_2=b_1^{-1}$. Then
\begin{equation*}
 Q_{n-1}(\zeta):=f_n\,,\quad\forall n\in\mathbb{N}\,.
\end{equation*}
$Q_n(\zeta)$ is a polynomial of degree $n-1$.

As pointed out in the introduction, $J$ has either deficiency indices
$(1,1)$ or $(0,0)$ \cite[Sec.\,1.2 Chap.\,4]{MR0184042} and
\cite[Cor.\,2.9]{MR1627806}. These cases correspond to the limit
circle and limit point case, respectively. In terms of the polynomials
of the first kind, $J$ has deficiency indices $(0,0)$ if for one
$\zeta\in\mathbb{C}\setminus\mathbb{R}$ the series in
(\ref{eq:generalized-eigenvector}) diverges.  In the limit circle case
(\ref{eq:generalized-eigenvector}) holds for every
$\zeta\in\mathbb{C}$ \cite[Thm.\,1.3.2]{MR0184042},
\cite[Thm.\,3]{MR1627806} and, therefore, $P(\zeta)$ is always in
$\Ker (J^*-\zeta I)$. Another peculiarity of the limit circle case is
that every self-adjoint extension of $J$ has purely discrete spectrum
\cite[Thm.\,4.11]{MR1627806}. Moreover, the resolvent of every
self-adjoint extension is a Hilbert-Schmidt operator
\cite[Lem.\,2.19]{MR1711536}.

In what follows we always consider $J$ to have deficiency indices
$(1,1)$. The behavior of the polynomials of the first kind determines
this class. There are various criteria for establishing whether a
Jacobi operator is symmetric but non-self-adjoint. These criteria may
be given in terms of the moments associated with the matrix, for
instance the criterion \cite[Prop.\,1.7]{MR1627806} due to
Krein. A criterion in terms of the matrix entries is the following
result which belongs to  Berezans{\cprime}ki{\u\i}
\cite[Chap.\,1]{MR0184042}, \cite[Thm.\,1.5 Chap.\,7]{MR0222718}.
\begin{proposition}
   Suppose that $\sup_{n\in\mathbb{N}}\abs{q_n}<\infty$ and that
  $\sum_{n=1}^\infty\frac{1}{b_n}<\infty$. If there is
  $N\in\mathbb{N}$ such that for $n>N$
  \begin{equation*}
    b_{n-1}b_{n+1}\le b_n^2\,,
  \end{equation*}
then the Jacobi operator whose associated matrix is (\ref{eq:jm-0}) is
in the limit circle case.
\end{proposition}
Jacobi operators in the limit circle case may be used to model
physical processes. For instance Krein's mechanical interpretation of
Stieltjes continued fractions \cite{MR0054078}, in which one has a
string carrying point masses with a certain distribution along the
string, is modeled by an eigenvalue equation of a Jacobi operator
\cite[Appendix]{MR0184042}. There are criteria in terms of the point
masses and their distribution \cite[Thm.\,0.4 Thm.\,0.5
Appendix]{MR0184042} for the corresponding Jacobi operator to be in
the limit circle case.

In this work, all self-adjoint extensions of $J$ are assumed
to be restrictions of $J^*$.  When dealing with all self-adjoint
extensions of $J$, including those which imply an extension of the
original Hilbert space, the self-adjoint restrictions of $J^*$ are
called von Neumann self-adjoint extensions of $J$
(cf. \cite[Appendix\,\,I]{MR1255973}, \cite[Sec.\,6]{MR1627806}).

There is also a well known result for $J$ in the limit circle case,
namely, that $J$ is simple \cite[Thm.\,4.2.4]{MR0184042}. In its turn
this imply that the eigenvalues of any self-adjoint extension of $J$
have multiplicity one \cite[Thm.3 Sec.\,81]{MR1255973}.

Let us now introduce a convenient way of parametrizing the
self-adjoint extensions of $J$ in the symmetric non-self-adjoint case.  We first
define the Wronskian associated with $J$ for any pair of sequences
$\varphi=\{\varphi_k\}_{k=1}^\infty$ and
$\psi=\{\psi_k\}_{k=1}^\infty$ in $l_2(\mathbb{N})$ as follows
\begin{equation*}
  W_k(\varphi,\psi):=b_k(\varphi_k\psi_{k+1}-\psi_k\varphi_{k+1})\,,
  \quad k\in\mathbb{N}\,.
\end{equation*}
Now,  consider the sequences $v(\tau)=\{v_k(\tau)\}_{k=1}^\infty$
such that, for $k\in\mathbb{N}$,
\begin{equation}
  \label{eq:boundary-sequence}
  v_k(\tau):=P_{k-1}(0)+\tau Q_{k-1}(0)\,,
  \quad \tau\in\mathbb{R}\,,
\end{equation}
and
\begin{equation}
  \label{eq:boundary-sequence-infty}
  v_k(\infty):=Q_{k-1}(0)\,.
\end{equation}

All the self-adjoint extensions $J(\tau)$ of the symmetric non-self-adjoint
operator $J$ are restrictions of $J^*$ to the set
\cite[Lem.\,2.20]{MR1711536}
\begin{equation}
  \label{eq:beta-extensions-domain}
 \mathcal{D}_\tau := \bigl\{f=\{f_k\}_{k=1}^\infty\in\dom (J^*):\,
 \lim_{n\to\infty}W_n\bigl(v(\tau),f\bigr)=0\bigr\}\,,\qquad
 \tau\in\mathbb{R}\cup\{\infty\}\,.
\end{equation}
Different values of $\tau$ imply different self-adjoint extensions, so
$J(\tau)$ is a self-adjoint extension of $J$ uniquely determined by
$\tau$ \cite[Lem.\,2.20]{MR1711536}. Observe that the domains
$\mathcal{D}_\tau$ are defined by a boundary condition at infinity
given by $\tau$. We also remark that given two sequences $\varphi$ and
$\psi$ in $\dom (J^*)$ the following limit always exists
\cite[Sec. 2.6]{MR1711536}
\begin{equation*}
  \lim_{n\to\infty}W_n(\varphi,\psi)=:W_\infty(\varphi,\psi)\,.
\end{equation*}
It follows from \cite[Thm.\,3]{MR1627806} that, in the limit circle
case, $P(\zeta)$ and $Q(\zeta)$ are in $\dom (J^*)$ for every
$\zeta\in\mathbb{C}$.

From what has just been said, one can consider the functions (see also
\cite[Sec.\,2.4 Chap.\,1, Sec.\,4.2 Chap.\,2]{MR0184042})
\begin{equation}
  \label{eq:b-d-definition}
\begin{split}
  W_\infty(P(0),P(\zeta))&=:D(\zeta)\,,\\
  W_\infty(Q(0),P(\zeta))&=:B(\zeta)\,.
\end{split}
\end{equation}
The notation for these limits has not been chosen arbitrarily; they
are the elements of the second row of the Nevanlinna matrix associated
with the matrix (\ref{eq:jm-0}) and they are usually denoted by these
letters \cite[Sec.\,4.2 Chap. 2]{MR0184042},
\cite[Eq.\,4.17]{MR1627806}.

It is well known that the functions $D(\zeta)$ and $B(\zeta)$ are
entire of at most minimal type of order one
\cite[Thm.\,2.4.3]{MR0184042}, \cite[Thm. 4.8]{MR1627806}, that is,
for each $\epsilon >0$ there exist constants
$C_1(\epsilon),\,C_2(\epsilon)$ such that
\begin{equation*}
  \abs{D(\zeta)}\le C_1(\epsilon)e^{\epsilon\abs{\zeta}}\,,\qquad
  \abs{B(\zeta)}\le C_2(\epsilon)e^{\epsilon\abs{\zeta}}\,.
\end{equation*}

If $P(\zeta)$ is in $\mathcal{D}_\tau$ the following holds
\begin{equation*}
  0=W_\infty(v(\tau),P(\zeta))=
\begin{cases}
D(\zeta)+\tau B(\zeta) & \quad\text{if}\quad \tau\in\mathbb{R}\\
B(\zeta) & \quad\text{if}\quad \tau=\infty\,.
\end{cases}
\end{equation*}
Thus, the zeros of the function
\begin{equation}
  \label{eq:R-function-def}
  \mathfrak{R}_\tau(\zeta):=
\begin{cases}
D(\zeta)+\tau B(\zeta)& \quad\text{if}\quad \tau\in\mathbb{R}\\
B(\zeta) & \quad\text{if}\quad \tau=\infty
\end{cases}
\end{equation}
constitute the spectrum of the self-adjoint extension $J(\tau)$ of
$J$.

A Jacobi matrix of the form (\ref{eq:jm-0}) determines, in a unique
way, the sequence $P(t)=\{P_{n-1}(t)\}_{n=1}^\infty$,
$t\in\mathbb{R}$. This sequence is orthonormal in any space
$L_2(\mathbb{R},d\rho)$, where $\rho$ is a solution of the moment
problem associated with the Jacobi matrix (\ref{eq:jm-0})
\cite[Sec.\,2.1 Chap.\,2]{MR0184042}. The elements of the sequence
$\{P_{n-1}(t)\}_{n=1}^\infty$ form a basis in $L_2(\mathbb{R},d\rho)$ if
$\rho$ is an N-extremal solution of the moment problem
\cite[Def. 2.3.3]{MR0184042} or, in other words, if $\rho$ can be
written as
\begin{equation}
  \label{eq:extremal-measure}
  \rho(t)=\langle E(t)e_1,e_1\rangle\,,\qquad  t\in\mathbb{R}\,,
\end{equation}
where $E(t)$ is the spectral resolution of the identity for some von
Neumann self-adjoint extension of the Jacobi operator $J$ associated
with (\ref{eq:jm-0}) \cite[Thm.\,2.3.3, Thm.\,4.1.4]{MR0184042}.

Let $\rho$ be given by (\ref{eq:extremal-measure}), then we can
consider the linear isometric operator $U$ which maps the canonical
basis $\{e_n\}_{n=1}^\infty$ in $l_2(\mathbb{N})$ into the orthonormal
basis $\{P_n(t)\}_{n=0}^\infty$ in $L_2(\mathbb{R},d\rho)$ as follows
\begin{equation}
  \label{eq:operator-u}
  Ue_n=P_{n-1}\,,\qquad n\in\mathbb{N}\,.
\end{equation}
By linearity, one extends $U$ to the span of $\{e_n\}_{n=1}^\infty$ and
by continuity, to all $l_2(\mathbb{N})$. Clearly, the range of $U$ is
all $L_2(\mathbb{R},d\rho)$. The Jacobi operator $J$ given by the
matrix (\ref{eq:jm-0}) is transformed by $U$ into the operator of
multiplication by the independent variable in $L_2(\mathbb{R},d\rho)$
if $J=J^*$, and into a symmetric restriction of the operator of
multiplication if $J\ne J^*$. Following the terminology used in
\cite{MR0190761}, we call the operator $UJU^{-1}$ in
$L_2(\mathbb{R},d\rho)$ the \emph{canonical representation} of $J$.

By virtue of the discreteness of $\Sp(J(\tau))$ in the limit circle
case (here and in the sequel, $\Sp(A)$ stands for the spectrum of
operator $A$), the function $\rho_\tau$ given by
(\ref{eq:extremal-measure}), with $E(t)$ being the resolution of the
identity of $J(\tau)$, can be written as follows
\begin{equation*}
  \rho_\tau(t)=\sum_{\lambda_k\le t}a(\lambda_k)^{-1}\,,
  \qquad \lambda_k\in\Sp(J(\tau))\,,
\end{equation*}
where the positive constant $a(\lambda_k)$ is the so-called
\emph{normalizing constant} of $J(\tau)$ corresponding to $\lambda_k$.
In the limit circle case it is easy to obtain the following formula
for the normalizing constants \cite[Sec.\,4.1 Chap\,3]{MR0184042},
\cite[Thm.\,4.11]{MR1627806}
\begin{equation}
  \label{eq:normalizing-constants-for-lambda}
  a(\lambda_k)=\norm{P(\lambda_k)}_{l_2(\mathbb{N})}^2\,,
  \qquad\lambda_k\in\Sp(J(\tau))\,.
\end{equation}
Formula (\ref{eq:normalizing-constants-for-lambda}), which gives the
jump of the spectral function at $\lambda_k$, also holds true in the
limit point case, when $\lambda_k$ is an eigenvalue of $J$
\cite[Thm.\,1.17 Chap.\,7]{MR0222718}.

It turns out that the spectral function $\rho_\tau$ uniquely
determines $J(\tau)$. Indeed, there are two ways of recovering the matrix from
the spectral function. One method, developed in \cite{MR1616422} (see
also \cite{MR1643529}), makes use of the asymptotic behaviour of the
Weyl $m$-function
\begin{equation*}
  m_\tau(\zeta):=\int_{\mathbb{R}}\frac{\rho_\tau(t)}{t-\zeta}
\end{equation*}
and the Ricatti equation \cite[Eq.\,2.15]{MR1616422},
\cite[Eq.\,2.23]{MR1643529},
\begin{equation}
  \label{eq:ricatti}
    b_n^2 m_\tau^{(n)}(\zeta)=
    q_n-\zeta-\frac{1}{m_\tau^{(n-1)}(\zeta)}\,,\quad n\in\mathbb{N}\,,
\end{equation}
where $m_\tau^{(n)}(\zeta)$ is the Weyl $m$-function of the
Jacobi operator associated with the matrix (\ref{eq:jm-0})
with the first $n$ columns and $n$ rows removed.

The other method for the reconstruction of the matrix is more
straightforward (see \cite[Sec.\,1.5 Chap.\,7 and, particularly,
Thm. 1.11]{MR0222718}).  The starting point is the sequence
$\{t^k\}_{k=0}^\infty$, $t\in\mathbb{R}$. From what we discussed
above, all the elements of the sequence $\{t^k\}_{k=0}^\infty$ are in
$L_2(\mathbb{R},d\rho_\tau)$ and one can apply, in this Hilbert space,
the Gram-Schmidt procedure of orthonormalization to the sequence
$\{t^k\}_{k=0}^\infty$.  One, thus, obtains a sequence of polynomials
$\{P_k(t)\}_{k=0}^\infty$ normalized and orthogonal in
$L_2(\mathbb{R},d\rho_\tau)$. These polynomials satisfy a three term
recurrence equation \cite[Sec.\,1.5 Chap.\,7]{MR0222718},
\cite[Sec.\,1]{MR1627806}
\begin{align}
\label{eq:favard-system1}
      tP_{k-1}(t) &= b_{k-1}P_{k-2}(t) +  q_k
      P_{k-1}(t) +  b_k P_k(t)\,,
\quad k \in \mathbb{N} \setminus \{1\}\,,\\
\label{eq:favard-system2}
 tP_0(t) &=  q_1 P_0(t) +  b_1 P_1(t)\,,
\end{align}
where all the coefficients $b_k$ ($k\in\mathbb{N}$) turn out to be
positive and $q_k$ ($k\in\mathbb{N}$) are real numbers.  The system
(\ref{eq:favard-system1}) and (\ref{eq:favard-system2}) defines a
matrix which is the matrix representation of $J$.

After obtaining the matrix associated with $J$, if it turns out to be
non-self-adjoint, one can easily obtain the boundary condition at
infinity which defines the domain of $J(\tau)$. The recipe is based on
the fact that the spectra of different self-adjoint extensions are
disjoint \cite[Sec.\,2.4 Chap.\,4]{MR0184042}. Take an eigenvalue,
$\lambda$, of $J(\tau)$, i.\,e., $\lambda$ is a point of discontinuity of
$\rho_\tau$ or a pole of $m_\tau$. Since the corresponding eigenvector
$P(\lambda)=\{P_{k-1}(\lambda)\}_{k=1}^\infty$ is in $\dom (J(\tau))$, it
must be that
\begin{equation*}
  W_\infty\bigl(v(\tau),P(\lambda)\bigr)=0\,.
\end{equation*}
This implies that either
$W_\infty\bigl(Q(0),P(\lambda)\bigr)=0$,
which means that $\tau=\infty$, or
\begin{equation*}
  \tau=-\frac
  {W_\infty\bigl(P(0),P(\lambda)\bigr)}
  {W_\infty\bigl(Q(0),P(\lambda)\bigr)}
  \,.
\end{equation*}

\noindent\textbf{Notation} We conclude this section with a remark on
the notation.  The elements of the unbounded set $\sigma(J(\tau))$,
$\tau\in\mathbb{R}\cup\infty$, may be enumerated in different
ways. Let $\sigma(J(\tau))=\{\lambda_k\}_{k\in K}$, where $K$ is a
countable set through which the subscript $k$ runs. If
$\sigma(J(\tau))$ is either bounded from above or below, one may take
$K=\mathbb{N}$. If $\sigma(J(\tau))$ is unbounded below and above, one
may set $K=\mathbb{Z}$. Of course, other choices of $K$ are
possible. Since the particular choice of $K$ is not important in our
formulae, we shall drop $K$ from the notation and simple write
$\{\lambda_k\}_k$. All our formulae will be written so that they are
\emph{independent} of the way the elements of a sequence are
enumerated, so our convention for denoting sequences should not lead
to misunderstanding. Similarly, we write $\sum_k y_k$ instead of
$\sum_{k\in K} y_k$, and the convergence of the series to a number $c$
means that for any sequence of sets $\{K_j\}_{j=1}^\infty$, with
$K_j\subset K_{j+1}\subset K$, such that $\bigcup_{j=1}^\infty K_j=K$,
the sequence $\{\sum_{k\in K_j}y_k\}_{j=1}^\infty$ tends to $c$
whenever $j\to\infty$.

\section{Unique reconstruction of the matrix}
\label{sec:recovering}
In this section we show that, given the spectra of two different
self-adjoint extensions $J({\tau_1})$, $J({\tau_2})$ of the Jacobi operator $J$
in the limit circle case, one can always recover the matrix, being the
matrix representation of $J$ with respect to the canonical basis in
$l_2(\mathbb{N})$, and the two parameters ${\tau_1}$, ${\tau_2}$ that define
the self-adjoint extensions. It has already been announced
\cite[Thm.\,1]{MR1045318} that, when ${\tau_1},{\tau_2}\in\mathbb{R}$ and
${\tau_1}\ne {\tau_2}$, the spectra $\Sp (J({\tau_1}))$ and $\Sp (J({\tau_2}))$ uniquely
determine the matrix of $J$ and the numbers ${\tau_1}$ and ${\tau_2}$. A similar
result, but in a more general setting can be found in
\cite[Thm.\,7]{MR0190761}.

Consider the following expression which follows from the
Christoffel-Darboux formula \cite[Eq.\,1.17]{MR0184042}:
\begin{equation*}
  \sum_{k=0}^{n-1}
  P_k^2(\zeta)=b_n\left(P_{n-1}(\zeta)P_n'(\zeta)-
  P_n(\zeta)P_{n-1}'(\zeta)\right)=W_n(P(\zeta),P'(\zeta))\,.
\end{equation*}
It is easy to verify, taking into account the analogue of the
Liouville-Ostrogradskii formula \cite[Eq.\,1.15]{MR0184042}, that
\begin{equation*}
\begin{split}
  W_n(P(\zeta),P'(\zeta))&=W_n(P(0),P(\zeta))W_n(Q(0),P'(\zeta))\\
&-W_n(Q(0),P(\zeta))W_n(P(0),P'(\zeta))\,.
\end{split}
\end{equation*}
Thus,
\begin{equation*}
  \begin{split}
   \sum_{k=0}^\infty
  P_k^2(\zeta)&=W_\infty(P(\zeta),P'(\zeta))\\
  &=D(\zeta)B'(\zeta)-B(\zeta)
 D'(\zeta)\,.
\end{split}
\end{equation*}
Indeed, due to the uniform convergence of the limits in
(\ref{eq:b-d-definition}) \cite[Sec.\,4.2 Chap.\,2]{MR0184042}, the
following is valid
\begin{equation*}
 \begin{split}
  B'(\zeta)&=W_\infty(Q(0),P'(\zeta))\\
  D'(\zeta)&=W_\infty(P(0),P'(\zeta))\,.
\end{split}
\end{equation*}
Now, a straightforward computation yields (${\tau_1},{\tau_2}\in\mathbb{R}$,
${\tau_1}\ne {\tau_2}$)
\begin{equation*}
  \mathfrak{R}_{\tau_1}(\zeta)\mathfrak{R}_{\tau_2}'(\zeta)-\mathfrak{R}_{\tau_1}'(\zeta)
  \mathfrak{R}_{\tau_2}(\zeta)=({\tau_2}-{\tau_1})\left[D(\zeta)B'(\zeta)
    -B(\zeta)D'(\zeta)\right]\,.
\end{equation*}
On the other hand one clearly has
\begin{equation*}
  \mathfrak{R}_{\tau_1}(\zeta)\mathfrak{R}_\infty'(\zeta)-\mathfrak{R}_{\tau_1}'(\zeta)
  \mathfrak{R}_\infty(\zeta)=D(\zeta)B'(\zeta) - B(\zeta)D'(\zeta)\,,
  \quad {\tau_1}\in\mathbb{R}\,.
\end{equation*}
Hence,
\begin{equation}
  \label{eq:normalizing}
  a(\zeta):=\sum_{k=0}^\infty
  P_k^2(\zeta)=
\begin{cases}
  \displaystyle\frac{\mathfrak{R}_{\tau_1}(\zeta)
    \mathfrak{R}_{\tau_2}'(\zeta)-\mathfrak{R}_{\tau_1}'(\zeta)
    \mathfrak{R}_{\tau_2}(\zeta)}{{\tau_2}-{\tau_1}} & {\tau_1}\ne {\tau_2}\,,\quad
  {\tau_1},{\tau_2}\in\mathbb{R}\\[4mm]
  \mathfrak{R}_{\tau_1}(\zeta)\mathfrak{R}_\infty'(\zeta)-\mathfrak{R}_{\tau_1}'(\zeta)
  \mathfrak{R}_\infty(\zeta) & {\tau_1}\in\mathbb{R}\,.
\end{cases}
\end{equation}
It follows from (\ref{eq:normalizing-constants-for-lambda}) that the
values of the function $a(\zeta)$ evaluated at the points of the
spectrum of some self-adjoint extension of $J$ are the corresponding
normalizing constants of that extension.

The analogue of (\ref{eq:normalizing}) with ${\tau_1}\ne {\tau_2}$ and
${\tau_1},{\tau_2}\in\mathbb{R}$, for the Schr\"{o}dinger operator in
$L_2(0,\infty)$ being in the limit circle case is
\cite[Eq.\,1.20]{MR997788}. Formula \cite[Eq.\,1.20]{MR997788} plays a
central r\^{o}le in proving the unique reconstruction theorem for that
operator \cite[Thm.\,1.1]{MR997788}. The discrete counterpart of
\cite[Thm.\,1.1]{MR997788} is \cite[Thm.\,1]{MR1045318}. It is worth
mentioning that the reconstruction technique we present below is also
based on (\ref{eq:normalizing}).

It is well known that the spectra of any two different self-adjoint
extensions of $J$ are disjoint \cite[Sec.\,2.4
Chap.\,4]{MR0184042}. One can easily conclude this from
(\ref{eq:normalizing}). Moreover, the following
assertion holds true.
\begin{proposition}
  \label{prop:interlacing}
  The eigenvalues of two different self-adjoint extensions of a Jacobi
  operator interlace, that is, there is only one eigenvalue of a
  self-adjoint extension between two eigenvalues of any other
  self-adjoint extension.
\end{proposition}
\begin{remark}
  One may arrive at
this assertion via rank-one perturbation theory, in particular by
recurring to the Aronzajn-Krein formula
\cite[Eq. 1.13]{simon1}. Nonetheless, we provide below a simple proof to illustrate the use of
(\ref{eq:normalizing}).  The proof of this statement for regular
simple symmetric operators can be found in \cite[Prop.\,3.4
Chap.\,1]{R1466698}.
\end{remark}
\begin{proof}
  The proof of this assertion follows from the expression
  (\ref{eq:normalizing}). It is similar to the proof of
  \cite[Thm.\,1.2.2]{MR0184042}.

  Note that (\ref{eq:R-function-def}) implies that the entire function
  $\mathfrak{R}_\tau(\zeta)$, $\tau\in\mathbb{R}\cup\{\infty\}$, is
  real, i.\,e., it takes real values when evaluated on the real
  line. Let $\lambda_k<\lambda_{k+1}$ be two neighboring eigenvalues
  of the self-adjoint extension $J({\tau_2})$ of $J$, with
  ${\tau_2}\in\mathbb{R}\cup\{\infty\}$. So $\lambda_k$, $\lambda_{k+1}$
  are zeros of $\mathfrak{R}_{\tau_2}$ and by (\ref{eq:normalizing}) these
  zeros are simple.  Since $\mathfrak{R}_{\tau_2}'(\lambda_k)$ and
  $\mathfrak{R}_{\tau_2}'(\lambda_{k+1})$ have different signs, it follows
  from (\ref{eq:normalizing}) that $\mathfrak{R}_{\tau_1}(\lambda_k)$ and
  $\mathfrak{R}_{\tau_1}(\lambda_{k+1})$ (${\tau_1}\in\mathbb{R}\cup\{\infty\},\,
  {\tau_1}\ne {\tau_2}$) have also opposite signs. From the continuity of
  $\mathfrak{R}_{\tau_1}$ on the interval $[\lambda_k,\lambda_{k+1}]$, there
  is at least one zero of $\mathfrak{R}_{\tau_1}$ in
  $(\lambda_k,\lambda_{k+1})$. Now, suppose that in this interval
  there is more than one zero of $\mathfrak{R}_{\tau_1}$, so one can take two
  neighboring zeros of $\mathfrak{R}_{\tau_1}$ in
  $(\lambda_k,\lambda_{k+1})$. By reproducing the argumentation above
  with ${\tau_1}$ and ${\tau_2}$ interchanged, one obtains that there is at least
  one zero of $\mathfrak{R}_{\tau_2}$ somewhere in
  $(\lambda_k,\lambda_{k+1})$. This contradicts the assumption that
  $\lambda_k$ and $\lambda_{k+1}$ are neighbors.\\
\end{proof}

The assertion of the following proposition is a well established fact
(see, for instance \cite[Thm.\,1]{MR0052004}). We, nevertheless,
provide the proof for the reader's convenience and because we
introduce in it notation for later use. Note that a non-constant
entire function of at most minimal type of order one must have zeros,
otherwise, by Weierstrass theorem on the representation of entire
functions by infinite products \cite[Thm.\,3 Chap.\,1]{MR589888}, it
would be a function of at least normal type.

Before stating the proposition we remind the definition of convergence
exponent of a sequence of complex numbers (see \cite[Sec.\,4
Chap.\,1]{MR589888}).  The convergence exponent $\rho_1$ of a sequence
$\{\nu_k\}_{k}$ of non-zero complex numbers accumulating only at
infinity is given by
  \begin{equation}
    \label{eq:convergence-exp}
    \rho_1:=\inf\,\left\{\gamma\in\mathbb{R}:\lim_{r\to\infty}\sum_{\abs{\nu_k}\le r}
      \frac{1}{\abs{\nu_k}^\gamma}<\infty\right\}\,.
  \end{equation}
We also remark that, as it is customary,
whenever we say that an infinite product is convergent we mean that at
most a finite number of factors may be zero and the partial product
formed by the non-vanishing factors tends to a number different from
zero \cite[Sec.\,2.2 Chap.\,5]{ahlfors}.
\begin{proposition}
\label{prop:entire-minimal-1-order-rep}
  Let $f(\zeta)$ be an entire function of at most minimal type of
  order one with an infinite number of zeros. Let the elements of the
  sequence $\{\nu_k\}_k$, which accumulate only at infinity, be the non-zero
  roots of $f$, where $\{\nu_k\}_k$ contains as many elements for each
  zero as its multiplicity. Assume that $m\in\mathbb{N}\cup\{0\}$ is the
  order of the zero of $f$ at the origin. Then there exists a complex
  constant $C$ such that
  \begin{equation}
    \label{eq:entire-minimal-1-order-representation}
    f(\zeta)=C\zeta^m\lim_{r\to\infty}
\prod_{\abs{\nu_k}\le r}
\left(1-\frac{\zeta}{\nu_k}\right)\,,
  \end{equation}
  where the limit converges uniformly on compacts of
  $\mathbb{C}$.
\end{proposition}
\begin{proof}
  The convergence exponent $\rho_1$ of the zeros of an arbitrary
  entire function does not exceed its order \cite[Thm.\,6
  Chap.\,1]{MR589888}. Then, for a function of at most minimal type of
  order one, $\rho_1\le 1$. According to Hadamard's theorem
  \cite[Thm.\,13 Chap.\,1]{MR589888}, the expansion of $f$ in an
  infinite product has either the form:
\begin{equation}
  \label{eq:hadamard1}
  f(\zeta)=\zeta^me^{a\zeta +b}\lim_{r\to\infty}\prod_{\abs{\nu_k}\le r}
  G\left(\frac{\zeta}{\nu_k};0\right)\,,\quad
  a,b\in\mathbb{C}
\end{equation}
if the limit
\begin{equation}
  \label{eq:limit-series}
  \lim_{r\to\infty}\sum_{\abs{\nu_k}\le r}
    \frac{1}{\abs{\nu_k}}
\end{equation}
converges, or
\begin{equation}
  \label{eq:hadamard2}
  f(\zeta)=\zeta^me^{c\zeta+d}\lim_{r\to\infty}\prod_{\abs{\nu_k}\le r}
  G\left(\frac{\zeta}{\nu_k};1\right)\,,\quad
  c,d\in\mathbb{C}
\end{equation}
if (\ref{eq:limit-series}) diverges. We have used here the Weierstrass
primary factors $G$ (for details see \cite[Sec.\,3
Chap.\,1]{MR589888}).  Let us suppose that the order is one and
(\ref{eq:limit-series}) diverges, then, in view of the fact that $f$
is of minimal type, by a theorem due to Lindel\"{o}f \cite[Thm.\,15\,a
Chap.\,1]{MR589888}, we have in particular that
\begin{equation*}
  \lim_{r\to\infty}
\sum_{\abs{\nu_k}\le r}\nu_k^{-1}=-c\,.
\end{equation*}
This implies the uniform convergence of the series $\lim_{r\to\infty}
\sum_{\abs{\nu_k}\le r}\frac{\zeta}{\nu_k}$ on compacts of
$\mathbb{C}$. In its turn, since $\rho_1= 1$, this yields that
$\lim_{r\to\infty}\prod_{\abs{\nu_k}\le
  r}\left(1-\frac{\zeta}{\nu_k}\right)$ is uniformly convergent on any
compact of $\mathbb{C}$. Therefore,
\begin{equation*}
  \lim_{r\to\infty}\prod_{\abs{\nu_k}\le r}
  G\left(\frac{\zeta}{\nu_k};1\right)=e^{-c\zeta}\lim_{r\to\infty}
\prod_{\abs{\nu_k}\le r}
 \left(1-\frac{\zeta}{\nu_k}\right)\,.
\end{equation*}
Thus, (\ref{eq:hadamard2}) can be written as
(\ref{eq:entire-minimal-1-order-representation}).

Suppose now that the limit (\ref{eq:limit-series}) converges. If the
order of the function is less than one, then, by \cite[Thm.\,13
Chap.\,1]{MR589888}, one may write (\ref{eq:hadamard1}) as
(\ref{eq:entire-minimal-1-order-representation}). If the order of the
function is one, by \cite[Thm. 12, Thm.\,15\,b
Chap.\,1]{MR589888}, one concludes again that (\ref{eq:hadamard1}) can
be written as (\ref{eq:entire-minimal-1-order-representation})
(cf. Thm 15 in the Russian version of \cite{MR589888} or,
alternatively, \cite[Lect. 5]{MR1400006}).
\\
\end{proof}

Let $\{\lambda_n(\tau)\}_n$ be the eigenvalues of $J(\tau)$. In view of
the fact that $\mathfrak{R}_\tau(\zeta)$ is an entire function of at most
minimal type of order one, by
Proposition~\ref{prop:entire-minimal-1-order-rep}, one can always
write
\begin{equation}
\label{eq:R-expansion}
  \mathfrak{R}_\tau(\zeta)=C_\tau\zeta^{\delta_{\tau}}\lim_{r\to\infty}
\prod_{0<\abs{\lambda_k(\tau)}\le r}
\left(1-\frac{\zeta}{\lambda_k(\tau)}\right)\,,\qquad
\tau\in\mathbb{R}\cup\{\infty\}\,,
\end{equation}
where $C_\tau\in\mathbb{R}\setminus\{0\}$ and $\delta_{\tau}$ is the Kronecker
delta, i.\,e., $\delta_\tau=1$ if $\tau=0$, and $\delta_\tau=0$
otherwise.  The limits in (\ref{eq:R-expansion}) converge uniformly on
compacts of $\mathbb{C}$. Note that when $\tau=0$ we have naturally
excluded $\lambda_k(0)=0$ from the infinite product.

When writing (\ref{eq:R-expansion}), we have taken into account, on
the one hand, that $\mathfrak{R}_0(0)=0$, which follows from
(\ref{eq:R-function-def}) and the definition of the function $D$, and
on the other, that different self-adjoint extensions have disjoint
spectra (see Section~\ref{sec:preliminaries}).

Now, let us consider the following expressions derived from
the Green's formula \cite[Eqs.\,1.23, 2.28]{MR0184042}
\begin{equation}
\label{eq:p}
  D(\zeta)=\zeta\sum_{k=0}^\infty P_k(0)P_k(\zeta)\,,
\end{equation}
\begin{equation}
\label{eq:q}
  B(\zeta)=-1+\zeta\sum_{k=0}^\infty Q_k(0)P_k(\zeta)\,.
\end{equation}
Again we verify from (\ref{eq:p}) that $D(0)=0$, while from
(\ref{eq:q}) we have $B(0)=-1$. Therefore $\mathfrak{R}_\tau(0)=-\tau$ for
every $\tau\in\mathbb{R}$, and $\mathfrak{R}_\infty(0)=-1$. Thus,
$C_\tau=-\tau$ provided that $\tau\in\mathbb{R}$ and $\tau\ne 0$, and
$C_\infty=-1$.

To simplify the writing of some of the formulae below, let us introduce
$R_\tau(\zeta):=\frac{\mathfrak{R}_\tau(\zeta)}{C_\tau}$, that is,
\begin{equation}
   \label{eq:R-without-constants}
R_\tau(\zeta):=\zeta^{\delta_{\tau}}\lim_{r\to\infty}
\prod_{0<\abs{\lambda_k(\tau)}\le r}
\left(1-\frac{\zeta}{\lambda_k(\tau)}\right)\,,\qquad
\tau\in\mathbb{R}\cup\{\infty\}\,,
\end{equation}
where $\delta_\tau$ is defined as in (\ref{eq:R-expansion}).

Due to the uniform convergence of the expression
  \begin{equation*}
    \frac{d}{d\zeta}\left[\prod_{0<\abs{\lambda_k(\tau)}\le
  r}\left(1-\frac{\zeta}{\lambda_k(\tau)}\right)\right]\,,\quad\text{ as }\quad
   r\to\infty\,,
  \end{equation*}
one has
\begin{equation}
  \label{eq:R-derivative}
  R_\tau'(\lambda_j(\tau))=
\begin{cases}
  -[\lambda_j(\tau)]^{\delta_\tau-1}\lim\limits_{r\to\infty}
\prod\limits_{\substack{0<\abs{\lambda_k(\tau)}\le r\\ k\ne j}}
\left(1-\frac{\lambda_j(\tau)}{\lambda_k(\tau)}\right) &
\lambda_j(\tau)\ne 0\\
1& \lambda_j(\tau)= 0
\end{cases}
\end{equation}
By (\ref{eq:normalizing}) and  (\ref{eq:R-derivative}), one obtains
\begin{equation}
  \label{eq:C-0}
  C_0=a(0).
\end{equation}

\begin{theorem}
  \label{thm:recovering}
  Let ${\tau_1},{\tau_2}\in\mathbb{R}\cup\{\infty\}$ with ${\tau_1}\ne {\tau_2}$.  The
  spectra $\{\lambda_k({\tau_1})\}_k$, $\{\lambda_k({\tau_2})\}_k$ of two
  different self-adjoint extensions $J({\tau_1})$, $J({\tau_2})$ of a Jacobi
  operator $J$ in the limit circle case uniquely determine
  the matrix associated with $J$, and the numbers ${\tau_1}$ and ${\tau_2}$.
\end{theorem}
\begin{proof}
  For definiteness assume that ${\tau_1}\ne 0$, in other words that the
  sequence $\{\lambda_k({\tau_1})\}_k$ does not contain
  any zero element.

By (\ref{eq:normalizing}), we have
  \begin{equation}
    \label{eq:normalizing-constants}
    a(\lambda_k({\tau_1}))=MR_{\tau_2}(\lambda_k({\tau_1}))
    R_{\tau_1}'(\lambda_k({\tau_1}))\,,
  \end{equation}
where
\begin{equation}
  \label{eq:constant-m}
  M=
  \begin{cases}
    \frac{{\tau_1}{\tau_2}}{{\tau_1}-{\tau_2}} & \text{if}\quad
    {\tau_1},{\tau_2}\in\mathbb{R}\setminus\{0\}\\
    {\tau_2} & \text{if}\quad {\tau_1}=\infty\,,\quad {\tau_2}\ne 0\\
    -{\tau_1} & \text{if}\quad {\tau_2}=\infty\\
    -C_0 & \text{if}\quad {\tau_2}=0
  \end{cases}
\end{equation}
Now, since $\{a(\lambda_k({\tau_1}))\}_k$ are the normalizing constants of
$J({\tau_1})$ we must have
\begin{equation*}
  1=\sum_k\frac{1}{a(\lambda_k({\tau_1}))}=\frac{1}{M}
  \sum_k\frac{1}{R_{\tau_2}(\lambda_k({\tau_1}))
    R_{\tau_1}'(\lambda_k({\tau_1}))}\,.
\end{equation*}
Therefore
\begin{equation}
  \label{eq:calculation-M}
  M=\sum_k\frac{1}{R_{\tau_2}(\lambda_k({\tau_1}))
    R_{\tau_1}'(\lambda_k({\tau_1}))}\,.
\end{equation}
Thus, $M$ is completely determined by the sequences
$\{\lambda_k({\tau_2})\}_k$ and $\{\lambda_k({\tau_1})\}_k$.  Inserting the obtained
value of $M$ into (\ref{eq:normalizing-constants}) one obtains the
normalizing constants. Having the normalizing constants allows us to
construct the spectral measure for $J({\tau_1})$. Then, by standard methods
(see Section~\ref{sec:preliminaries}), one reconstructs the matrix
associated with $J$ and the boundary condition at infinity ${\tau_1}$. From
the value of $M$ and ${\tau_1}$ one obtains ${\tau_2}$, by using the first
three cases in (\ref{eq:constant-m}). When $0\in\{\lambda_k({\tau_2})\}_k$,
one does not use (\ref{eq:constant-m}), since it is already known that ${\tau_2}=0$.
\\
\end{proof}
\begin{remark}
  Note that the proof of Theorem~\ref{thm:recovering} gives a
  reconstruction method of the Jacobi matrix. Although mentioned
  earlier, we also remark here that the assertion of
  Theorem~\ref{thm:recovering}, for the case of ${\tau_1},{\tau_2}\in\mathbb{R}$,
  was announced without proof in \cite[Thm.\,1]{MR1045318}.
\end{remark}

\section{Necessary and sufficient conditions}
\label{sec:n-s-conditions}
In this section we give a complete characterization of our
two-spectra inverse problem. We remind the reader about the remark on
the notation at the end of Section~\ref{sec:preliminaries}.

First we prove the following simple proposition related to the
converse of Proposition~\ref{prop:entire-minimal-1-order-rep}.
\begin{proposition}
  \label{prop:prod-to-entire-mtoo}
  Let $\{\nu_k\}_k$ be an infinite sequence of non-vanishing complex
  numbers accumulating only at $\infty$, and whose
  convergence exponent $\rho_1$ does not exceed one. Suppose that the
  infinite product
\begin{equation}
\label{eq:infinite-prod}
  \lim_{r\to\infty}
\prod_{\abs{\nu_k}\le r}
\left(1-\frac{\zeta}{\nu_k}\right)
\end{equation}
converges uniformly on any compact of $\mathbb{C}$. Then this product
is an entire function of at most minimal type of
order one if either (\ref{eq:limit-series}) converges or if
(\ref{eq:limit-series}) diverges but the following holds
\begin{equation}
  \label{eq:density}
  \lim_{r\to\infty}\frac{n(r)}{r}=0\,,
\end{equation}
where $n(r)$ is the number of elements of $\{\nu_k\}_k$ in the circle
$\abs{\zeta}<r$.
\end{proposition}
\begin{proof}
  Clearly, by the conditions of the theorem, one can express
  (\ref{eq:infinite-prod}) in terms of canonical products
  \cite[Sec.\,3 Chap.\,1]{MR589888} either in the form
\begin{equation}
    \label{eq:first-case}
     \lim_{r\to\infty}
\prod_{\abs{\nu_k}\le r}
\left(1-\frac{\zeta}{\nu_k}\right)=\lim_{r\to\infty}\prod_{\abs{\nu_k}\le r}
  G\left(\frac{\zeta}{\nu_k};0\right)
\end{equation}
whenever (\ref{eq:limit-series}) converges, or in the form
\begin{equation}
  \label{eq:second-case}
   \lim_{r\to\infty}
\prod_{\abs{\nu_k}\le r}
\left(1-\frac{\zeta}{\nu_k}\right)=e^{c\zeta}\lim_{r\to\infty}\prod_{\abs{\nu_k}\le r}
  G\left(\frac{\zeta}{\nu_k};1\right)
\end{equation}
otherwise, where
\begin{equation}
  \label{eq:zero-distribution}
 \lim_{r\to\infty} \sum_{\abs{\nu_k}\le
  r}\nu_k^{-1}=-c\,.
\end{equation}
In the case (\ref{eq:first-case}), in which the genus of the product is
less than the convergence exponent of $\{\nu_k\}_k$, it is clear that
(\ref{eq:infinite-prod}) does not grow faster than an entire function
of minimal type of order one. Indeed, by \cite[Thm.\,7
Chap.\,1]{MR589888}, the order of a canonical product is equal to the
convergence exponent, so when $\rho_1<1$ the assertion is
obvious. For $\rho_1=1$ the statement follows from \cite[Thm.\,15\,b
Chap.\,1]{MR589888}.

If we have the representation (\ref{eq:second-case}), then
$\rho_1=1$. By \cite[Thm.\,7 Chap.\,1]{MR589888}, the canonical
product has order one. Since the product of functions of the same
order is of that same order, the order of (\ref{eq:infinite-prod}) is
one. Then, the assertion follows from \cite[Thm.\,15\,a
Chap.\,1]{MR589888} due to (\ref{eq:density}) and
(\ref{eq:zero-distribution}).\\
\end{proof}
Before passing on to the main results of this section, we establish an
auxiliary result which is related to part of the proof of Theorem 1 in
the Addenda and Problems of \cite[Chap.\,4]{MR0184042}.
\begin{lemma}
\label{lem:auxiliary}
  Consider an infinite real sequence $\{\kappa_j\}_j$ and a sequence
  $\{\alpha_j\}_j$ of positive numbers such that
  \begin{equation*}
    \sum_j\frac{\kappa_j^{2m}}{\alpha_j}<\infty\quad
    \text{ for all}\ m=0,1,\dots
\end{equation*}
Let $\mathfrak{F}$ be an entire function of at most minimal type of
order one whose zeros,
$\{\kappa_j\}_j$, are simple, and such that
\begin{equation}
  \label{eq:unboundedness-along-imaginary}
  \abs{\mathfrak{F}(\I t)}\to\infty\ \text{ as }\
  t\to\pm\infty\,,\quad t\in\mathbb{R}\,.
\end{equation}
If
\begin{equation*}
  \sum_j\frac{\alpha_j}{(1+\kappa_j^2)[\mathfrak{F}'(\kappa_j)]^2}<\infty\,,
\end{equation*}
then
\begin{equation}
  \label{eq:lemma-absolute-convergence}
  \sum_j\frac{\kappa_j^m}{\mathfrak{F}'(\kappa_j)}
\end{equation}
is absolutely convergent for $m=0,1,\dots$, and the absolutely
convergent expansion
\begin{equation*}
  \frac{1}{\mathfrak{F}(\zeta)}=
  \sum_j\frac{1}{\mathfrak{F}'(\kappa_j)(\zeta-\kappa_j)}
\end{equation*}
holds true for all $\zeta\in\mathbb{C}\setminus \{\kappa_j\}_j$.
\end{lemma}
\begin{proof}
  The absolutely convergence of (\ref{eq:lemma-absolute-convergence})
  follows from
  \begin{equation*}
    \begin{split}
      \sum_j\abs{\frac{\kappa_j^m}{\mathfrak{F}'(\kappa_j)}}&=
      \sum_j\abs{\frac{\sqrt{\alpha_j}}{\sqrt{1+\kappa_j^2}\mathfrak{F}'(\kappa_j)}}
      \abs{\frac{\kappa_j^m\sqrt{1+\kappa_j^2}}{\sqrt{\alpha_j}}}\\
      &\le\sqrt{\sum_j\frac{\alpha_j}{(1+\kappa_j^2)[\mathfrak{F}'(\kappa_j)]^2}}
      \sqrt{\sum_j\frac{\kappa_j^{2m}+\kappa_j^{2m+2}}{\alpha_j}}<\infty
    \end{split}
  \end{equation*}
Construct the function
\begin{equation*}
  h(\zeta):=1-\mathfrak{F}(\zeta)\sum_j\frac{1}
  {\mathfrak{F}'(\kappa_j)(\zeta-\kappa_j)}\,,
\end{equation*}
where the series is absolutely convergent in compact subsets of
$\mathbb{C}\setminus\{\kappa_j\}_j$ because of
(\ref{eq:lemma-absolute-convergence}). Clearly, $h(\kappa_j)=0$ for
any $j$. Moreover, it turns out that $h$ is an entire function of at
most minimal type of order one. To show this, first consider the case
when $\sum_j\abs{\kappa_j}^{-1}<\infty$. Here, by what we have
discussed in the proof of
Proposition~\ref{prop:entire-minimal-1-order-rep}, the function
$\mathfrak{F}(\zeta)/(\zeta-\kappa_j)$ can be expressed by a
  canonical product of genus zero. It follows from \cite[Lem.\,3
Chap.\,1]{MR589888} (see also the proof of \cite[Thm.\,4
Chap.\,1]{MR589888}) that, on the one hand,
\begin{equation*}
  \max_{\abs{\zeta}=r}\abs{\frac{\mathfrak{F}(\zeta)}{(\zeta-\kappa_j)}}<
  \exp\left(C(\alpha)r^\alpha\right)\,,\quad \rho_1<\alpha<1\,,
\end{equation*}
for any $r>0$ provided that $\rho_1<1$. If, on the other hand,
$\rho_1=1$, then for any $\epsilon>0$, there exists $R_0>0$ such that
\begin{equation}
  \label{eq:uniform-j-estimation}
  \max_{\abs{\zeta}=r}\abs{\frac{\mathfrak{F}(\zeta)}{(\zeta-\kappa_j)}}<
\exp\left(\epsilon
    r\right)
\end{equation}
for all $r>R_0$. Hence, in any case, we have the uniform, with respect
to $j$, asymptotic estimation (\ref{eq:uniform-j-estimation}) when
$\sum_j\abs{\kappa_j}^{-1}<\infty$.

Suppose now that $\sum_j\abs{\kappa_j}^{-1}=\infty$. In this case, as
was shown in the proof of Proposition~\ref{prop:entire-minimal-1-order-rep},
\begin{equation*}
  \frac{\mathfrak{F}(\zeta)}{(\zeta-\kappa_j)}=-\frac{1}{\kappa_j}\zeta^me^{(c+\kappa_j^{-1})\zeta+d}
  \lim_{r\to\infty}
\prod_{\substack{\abs{\kappa_k}\le r\\ k\ne j}}G\left(\frac{\zeta}{\kappa_k};1\right)\,,
\end{equation*}
where
\begin{equation}
  \label{eq:limit-to--c}
  \lim_{r\to\infty}\sum_{\abs{\kappa_k}\le r}\kappa_k^{-1}=-c\,.
\end{equation}
On the basis of the estimates found in the proof of \cite[Thm.\,15
Chap.\,1]{MR589888} (see in particular the inequality next to
\cite[Eq.\,1.43]{MR589888}), one can find $R_1$, independent of $j$,
such that
\begin{equation*}
  \max_{\abs{\zeta}=r}\abs{\frac{\mathfrak{F}(\zeta)}{(\zeta-\kappa_j)}}<
  \exp\left[r\left(\abs{c+\sum_{\abs{\kappa_k}\le r}\kappa_k^{-1}} +
C\left(\limsup_{r\to\infty}\frac{n(r)}{r}+\epsilon\right)+
O\left(\frac{1}{r}\right)\right)\right]
\end{equation*}
for all $r>R_1$ and $\epsilon>0$ (see the definition of $n(r)$ in the
statement of Proposition~\ref{prop:prod-to-entire-mtoo}). Note that if
$\abs{\kappa_j}\le r$, then the above inequality follows directly from
the inequality next to \cite[Eq.\,1.43]{MR589888}. If
$\abs{\kappa_j}> r$, the same inequality holds due to
\begin{equation*}
  \abs{c+\kappa_j^{-1}+\sum_{\abs{\kappa_k}\le r}\kappa_k^{-1}}\le
\abs{c+\sum_{\abs{\kappa_k}\le r}\kappa_k^{-1}}+\frac{1}{r}\,.
\end{equation*}
Since $\mathfrak{F}$ does not grow faster than a function of minimal
type of order one, by \cite[Thm.\,15\,a Chap.\,1]{MR589888}, one again
verifies that, for any $\epsilon>0$, (\ref{eq:uniform-j-estimation})
holds for all $r$ greater than a certain $R_2$ depending only on the
velocity of convergence in the limits (\ref{eq:limit-to--c}) and
(\ref{eq:density}).

Thus, one concludes that, for any $\epsilon>0$, there is $R>0$ such
that
\begin{equation*}
  \max_{\abs{\zeta}=r}\abs{\mathfrak{F}(\zeta)\sum_j\frac{1}
  {\mathfrak{F}'(\kappa_j)(\zeta-\kappa_j)}}\le
\sum_j\frac{1}{\abs{\mathfrak{F}'(\kappa_j)}}
\max_{\abs{\zeta}=r}\abs{\frac{\mathfrak{F}(\zeta)}{(\zeta-\kappa_j)}}<\exp(\epsilon
r)
\end{equation*}
for all $r>R$, which shows that $h$ is an entire function of at
most minimal type of order one.

Now, the function $h/\mathfrak{F}$ is also an entire function of at
most minimal type of order one \cite[Cor. Sec.\,9  Chap.\,1]{MR589888}.
By the hypothesis (\ref{eq:unboundedness-along-imaginary}),
\begin{equation*}
  \lim_{\substack{t\to\pm\infty \\ t\in\mathbb{R}}}
\frac{h(\I t)}{\mathfrak{F}(\I t)}=0\,,
\end{equation*}
which implies that $h/\mathfrak{F}\equiv 0$ (see Corollary of
\cite[Sec.\,14 Chap.\,1]{MR589888}).\\
\end{proof}

\begin{theorem}
  \label{thm:necessary-sufficient}
  Let $\{\lambda_k\}_k$ and  $\{\mu_k\}_k$ be two infinite sequences
  of real numbers such that
  \begin{enumerate}[a)]
  \item $\{\lambda_k\}_k\cap\{\mu_k\}_k=\emptyset$. For definiteness we assume that
    $0\not\in\{\lambda_k\}_k$
    \item the sequences accumulate only at the point at infinity.
    \item $\lambda_k\ne\lambda_j$, $\mu_k\ne\mu_j$ for $k\ne j$.
  \end{enumerate}
  Then there exist unique ${\tau_1},{\tau_2}\in\mathbb{R}\cup\{\infty\}$, with
  ${\tau_1}\ne0$, ${\tau_1}\ne {\tau_2}$, and a unique Jacobi operator $J\ne J^*$
  such that $\{\lambda_k\}_k=\Sp (J({\tau_1}))$ and $\{\mu_k\}_k=\Sp
  (J({\tau_2}))$ if and only if the following conditions are satisfied.
\begin{enumerate}
\item The convergence exponents of the sequence $\{\lambda_k\}_k$, and
  of the non-zero elements of $\{\mu_k\}_k$ do not exceed one. Additionally, if
  \begin{equation*}
     \lim_{r\to\infty}\sum_{\abs{\lambda_k}\le r}
    \frac{1}{\abs{\lambda_k}}=\infty\,,\quad\text{ require that }\quad
  \lim_{r\to\infty}\frac{n_\lambda(r)}{r}=0\,,
  \end{equation*}
and if
\begin{equation*}
     \lim_{r\to\infty}\sum_{0<\abs{\mu_k}\le r}
    \frac{1}{\abs{\mu_k}}=\infty\,,\quad\text{ require that }\quad
  \lim_{r\to\infty}\frac{n_\mu(r)}{r}=0\,,
\end{equation*}
where $n_\lambda(r)$ and $n_\mu(r)$ are the number of elements of
$\{\lambda_k\}_k$ and $\{\mu_k\}_k$, respectively, in the circle
$\abs{\zeta}<r$.
\label{nec-suf0}
\item The limits
  \begin{equation*}
\lim_{r\to\infty}
\prod_{\abs{\lambda_k}\le r}
\left(1-\frac{\zeta}{\lambda_k}\right)
\qquad
   \lim_{r\to\infty}
  \prod_{0<\abs{\mu_k}\le r}
  \left(1-\frac{\zeta}{\mu_k}\right)\,,
  \end{equation*}
converge uniformly on compact subsets of
$\mathbb{C}$, and they define the functions
\begin{equation}
  \label{eq:r-lambda-r-mu-1}
  \mathcal{R}_\lambda(\zeta):=
\lim_{r\to\infty}
\prod_{\abs{\lambda_k}\le r}
\left(1-\frac{\zeta}{\lambda_k}\right)
\end{equation}
\begin{equation}
  \label{eq:r-lambda-r-mu-2}
\mathcal{R}_\mu(\zeta):=\zeta^\delta\lim\limits_{r\to\infty}
  \prod\limits_{0<\abs{\mu_k}\le r}
  \left(1-\frac{\zeta}{\mu_k}\right)\,,
\end{equation}
where $\delta=1$ if $0\in\{\mu_k\}_k$, and $\delta=0$ otherwise.
\label{nec-suf1}
\item All numbers $\mathcal{R}_\mu(\lambda_j)\mathcal{R}_\lambda'(\lambda_j)$
have the same sign for all $j$. The same is true for the numbers
$\mathcal{R}_\lambda(\mu_j)\mathcal{R}_\mu'(\mu_j)$.
\label{nec-suf-constancy-sign}
\item For every $m=0,1,2,\dots$ the series below are convergent
  and the following equalities hold
  \begin{equation*}
\sum_j\frac{\lambda_j^m}{\mathcal{R}_\mu(\lambda_j)\mathcal{R}_\lambda'(\lambda_j)}=-
\sum_j\frac{\mu_j^m}{\mathcal{R}_\lambda(\mu_j)\mathcal{R}_\mu'(\mu_j)}
\end{equation*}\label{nec-suf-moments}
\item The series
  \begin{equation*}
\sum_j\frac{\mathcal{R}_\mu(\lambda_j)}{\mathcal{R}_\lambda'(\lambda_j)}
\quad\text{ and }\quad
\sum_j\frac{\mathcal{R}_\lambda(\mu_j)}{\mathcal{R}_\mu'(\mu_j)}
  \end{equation*}
diverge either to $-\infty$ or $+\infty$.\label{nec-suf-divergence}
\item The series
  \begin{equation*}
\sum_j\frac{\mathcal{R}_\mu(\lambda_j)}{(1+\lambda_j^2)\mathcal{R}_\lambda'(\lambda_j)}
\quad\text{ and }\quad
\sum_j\frac{\mathcal{R}_\lambda(\mu_j)}{(1+\mu_j^2)\mathcal{R}_\mu'(\mu_j)}
  \end{equation*}
are convergent.
\label{nec-suf3}
\end{enumerate}
\end{theorem}
\begin{proof}
  We begin by proving that if $\{\lambda_k\}_k$ and $\{\mu_k\}_k$ are,
  respectively, the spectra of the self-adjoint extensions $J({\tau_1})$ and
  $J({\tau_2})$ of a Jacobi operator $J$, then conditions
  \emph{\ref{nec-suf0}}--\emph{\ref{nec-suf3}} hold true.

  Since  $\{\lambda_k\}_k=\Sp (J({\tau_1}))$ and $\{\mu_k\}_k=\Sp(J({\tau_2}))$,
  the functions $\mathfrak{R}_{\tau_1}$ and $\mathfrak{R}_{\tau_2}$, given by
  (\ref{eq:R-function-def}), have the sequences $\{\lambda_k\}_k$ and
  $\{\mu_k\}_k$, respectively, as their sets of zeros. These functions
  do not grow faster than an entire function of minimal type of order
  one. By Proposition~\ref{prop:entire-minimal-1-order-rep} (see
  (\ref{eq:R-expansion})), the limits
  \begin{equation*}
    \lim_{r\to\infty}
    \prod_{\abs{\lambda_k}\le r}
    \left(1-\frac{\zeta}{\lambda_k}\right)\,,
    \qquad\lim_{r\to\infty}
    \prod_{0<\abs{\mu_k}\le r}
    \left(1-\frac{\zeta}{\mu_k}\right)
  \end{equation*}
  converge uniformly on compacts of $\mathbb{C}$. This is
  condition~\emph{\ref{nec-suf1}}. Moreover, by
  (\ref{eq:hadamard2}) and \cite[Thm.\,15\,a
  Chap.\,1]{MR589888}, condition \emph{\ref{nec-suf0}} holds.

  The functions $\mathcal{R}_\lambda$ and $\mathcal{R}_\mu$, given by
  (\ref{eq:r-lambda-r-mu-1}) and (\ref{eq:r-lambda-r-mu-2}), coincide
  with $R_\tau$, given by (\ref{eq:R-without-constants}), with
  $\tau={\tau_1}$ and $\tau={\tau_2}$, respectively. Thus
  (\ref{eq:normalizing-constants}) is rewritten as follows
\begin{equation}
  \label{eq:normalizing-necessary}
  a(\lambda_j)=M\mathcal{R}_\mu(\lambda_j)\mathcal{R}_\lambda'(\lambda_j)\,,
\end{equation}
where $M$ is given by (\ref{eq:constant-m}). Analogously,
\begin{equation}
 \label{eq:normalizing-necessary-mu}
 a(\mu_j)=-M\mathcal{R}_\lambda(\mu_j)\mathcal{R}_\mu'(\mu_j)\,.
\end{equation}

On the basis of the positiveness of the normalizing constants, from
(\ref{eq:normalizing-necessary}) and
(\ref{eq:normalizing-necessary-mu}), we obtain condition
\emph{\ref{nec-suf-constancy-sign}}.

From what we discussed in Section~\ref{sec:preliminaries} all the
moments exist for the spectral functions of $J({\tau_1})$ and $J({\tau_2})$, which
are, respectively,
\begin{equation}
  \label{eq:spectral-measures}
  \sum_{\lambda_k\le t}\frac{1}{a(\lambda_k)}\quad\text{ and }\quad
  \sum_{\mu_k\le t}\frac{1}{a(\mu_k)}\,.
\end{equation}
Hence the series in both sides of condition
\emph{\ref{nec-suf-moments}} are convergent for
$m\in\mathbb{N}\cup\{0\}$.  Moreover, the spectral functions
(\ref{eq:spectral-measures}) are solutions of the same moment problem
associated with $J$ (see the paragraph surrounding
(\ref{eq:extremal-measure})), therefore the equality of condition
\emph{\ref{nec-suf-moments}} holds.

Theorem 1 in the Addenda and Problems of \cite[Chap.\,4]{MR0184042}
tells us that
\begin{equation}
  \label{eq:akhiezer-conds}
  \sum_j\frac{a(\lambda_j)}
  {\left[\mathcal{R}_\lambda'(\lambda_j)\right]^2}=+\infty
\end{equation}
is a necessary condition for the sequences $\{\lambda_j\}_j$ and
$\{a(\lambda_j)\}_j$ to be the spectrum of $J({\tau_1})$ and its
corresponding normalizing constants. Thus, substituting
(\ref{eq:normalizing-necessary}) into
(\ref{eq:akhiezer-conds}), one establishes the divergence of the first
series in condition \emph{\ref{nec-suf-divergence}}. Similarly,
\begin{equation*}
   \sum_j\frac{a(\mu_j)}
  {\left[\mathcal{R}_\mu'(\mu_j)\right]^2}=+\infty
\end{equation*}
must hold, which, by (\ref{eq:normalizing-necessary-mu}),
implies the divergence of the second series in condition
\emph{\ref{nec-suf-divergence}}.

By the same theorem in \cite{MR0184042} mentioned above, and taking
into account 
(\ref{eq:normalizing-necessary}) and
(\ref{eq:normalizing-necessary-mu}), one obtains the convergence of the
series in condition \emph{\ref{nec-suf3}}.

Let us now prove that the conditions
\emph{\ref{nec-suf0}}--\emph{\ref{nec-suf3}} are sufficient. Using
condition \emph{\ref{nec-suf1}} and the convergence of the series in
the left hand side of condition \emph{\ref{nec-suf-moments}}  with
$m=0$, we define the
real constant
\begin{equation}
  \label{eq:definition-of-m-cal}
     \mathcal{M}:= \sum_j\frac{1}{\mathcal{R}_\mu(\lambda_j)\mathcal{R}_\lambda'(\lambda_j)}
\end{equation}
and the sequence of numbers
\begin{equation}
  \label{eq:sequence-def}
    a_j:=\mathcal{M}\mathcal{R}_\mu(\lambda_j)\mathcal{R}_\lambda'(\lambda_j)
\end{equation}
By condition \emph{\ref{nec-suf-constancy-sign}} and
(\ref{eq:definition-of-m-cal}), it follows that $a_j>0$ for all
$j$. Moreover, (\ref{eq:definition-of-m-cal}) and
(\ref{eq:sequence-def}) imply that $\sum_ja_j^{-1}=1$.

With the aid of the sequences $\{\lambda_k\}$ and $\{a_k\}$
define the function $\rho :\mathbb{R}\to\mathbb{R}_+$ as follows
\begin{equation}
  \label{eq:definition-sigma}
  \rho(t):=\sum_{\lambda_k\le t}a_k^{-1}.
\end{equation}
Consider the self-adjoint operator of multiplication $A_\rho$ by the
independent variable in $L_2(\mathbb{R},d\rho)$. We show below that
this operator is the canonical representation (see
Section~\ref{sec:preliminaries}) of a self-adjoint extension of a
Jacobi matrix in the limit circle case. The proof of this fact is
similar to the proof of Theorem 2 in Addenda and Problems of
\cite[Chap.\,4]{MR0184042}.  Note, however, that our conditions are
slightly different.

Consider a function $\theta_k(t)\in L_2(\mathbb{R},d\rho)$
such that
\begin{equation}
  \label{eq:definition-theta}
  \theta_k(\lambda_j)=\sqrt{a_k}\delta_{kj}\,.
\end{equation}
Clearly, $\theta_k(t)$ is the normalized eigenvector of $A_\rho$
corresponding to $\lambda_k$. Let $\varphi(t)\in
L_2(\mathbb{R},d\rho)$ be such that
\begin{equation}
  \label{eq:definition-phi}
  \langle\varphi,\theta_j\rangle_{L_2(\mathbb{R},d\rho)}=
\frac{\sqrt{a_j}}{(\lambda_j-\I)\mathcal{R}_\lambda'(\lambda_j)}\,.
\end{equation}
Taking into account
(\ref{eq:sequence-def}), it is clear that the convergence of the first
series in condition \emph{\ref{nec-suf3}} ensures that $\varphi(t)$ is indeed an
element of $L_2(\mathbb{R},d\rho)$. Define
\begin{equation}
  \label{eq:definition-domain}
  D:=\left\{\xi\in L_2(\mathbb{R},d\rho):\xi=(A_\rho+\I I)^{-1}\psi,
  \,\psi\in L_2(\mathbb{R},d\rho),\,\psi\perp\varphi\right\}\,.
\end{equation}
Since $D\subset\dom (A_\rho)$, we can consider the restriction of
$A_\rho$ to the linear set $D$. Let us show that this restriction
is a symmetric operator with deficiency indices $(1,1)$. First
we verify that $D$ is dense in
$L_2(\mathbb{R},d\rho)$. Suppose that a non-zero ${\eta}\in
L_2(\mathbb{R},d\rho)$ is orthogonal to $D$. This would imply
that there is a non-zero constant $C\in\mathbb{C}$ such that
\begin{equation*}
  {\eta}=C(A_\rho-\I I)\varphi\,.
\end{equation*}
Therefore,
\begin{equation*}
  \langle{\eta},\theta_j\rangle_{L_2(\mathbb{R},d\rho)}=
C\frac{\sqrt{a_j}}{\mathcal{R}_\lambda'(\lambda_j)}\,,
\end{equation*}
whence we easily conclude that ${\eta}\in L_2(\mathbb{R},d\rho)$
would contradict condition \emph{\ref{nec-suf-divergence}}.

Consider now the restriction of $A_\rho$ to the set $D$, denoted
henceforth by $A_\rho\upharpoonright_D$, and let us find the
dimension of $\ker ((A_\rho\upharpoonright_D)^*-\I I)$
which is characterized as the set of all $\omega\in
L_2(\mathbb{R},d\rho)$ for which the equation
\begin{equation*}
  \langle(A_\rho+\I I)\xi,\omega\rangle_{L_2(\mathbb{R},d\rho)}=0
\end{equation*}
is satisfied for any $\xi\in D$. It is not difficult to show
that any such $\omega$ can be written as follows
\begin{equation*}
  \omega=\widetilde{C}\varphi\,,\qquad 0\ne\widetilde{C}\in\mathbb{C}.
\end{equation*}
Hence $\dim\ker ((A_\rho\upharpoonright_D)^*-\I I)=1$. Analogously, it
can be shown that the dimension of $\ker
((A_\rho\upharpoonright_D)^*+\I I)$ also equals one. Indeed, if
$\omega\in\ker ((A_\rho\upharpoonright_D)^*+\I I)$ then, up to a
complex constant $\omega=(A_\rho-\I I)(A_\rho+\I I)^{-1}\varphi$\,.

Now we show that $A_\rho\upharpoonright_D$ is the
canonical representation of a Jacobi operator in the limit circle case
and $A_\rho$ is the canonical representation of a self-adjoint
extension of this Jacobi operator. We proceed stepwise.
\begin{enumerate}[I.]
\item We orthonormalize the sequence of functions
  $\{t^n\}_{n=0}^\infty$ with respect to the inner product of
  $L_2(\mathbb{R},d\rho)$. Note that condition
  \emph{\ref{nec-suf-moments}} guarantees that all elements of the
  sequence $\{t^n\}_{n=0}^\infty$ are in $L_2(\mathbb{R},d\rho)$. We
  obtain thus a sequence of polynomials $\{P_{n-1}(t)\}_{n=1}^\infty$
  which satisfy the three term recurrence equation
  (\ref{eq:favard-system1}) and (\ref{eq:favard-system2}),
where all the coefficients $b_k$ ($k\in\mathbb{N}$) turn out
to be positive and $q_k$ ($k\in\mathbb{N}$) are real numbers.

\item We verify that the polynomials are dense in
  $L_2(\mathbb{R},d\rho)$, so the sequence we have constructed is a
  basis in $L_2(\mathbb{R},d\rho)$. Note first that the function
  $\mathcal{R}_\lambda(\zeta)$ is entire of at most minimal type of order
  one. Indeed, this follows from
  Proposition~\ref{prop:prod-to-entire-mtoo}, in view of conditions
  \emph{\ref{nec-suf0}} and \emph{\ref{nec-suf1}}. Now, for any
  element $\lambda_{k_0}$ of the sequence $\{\lambda_k\}_k$, we
  clearly have
\begin{equation*}
  \abs{\mathcal{R}_\lambda(\I t)}\ge
  \abs{1+\frac{t^2}{\lambda_{k_0}^2}}\,,
  \qquad t\in\mathbb{R}\,.
\end{equation*}
This implies that $\mathcal{R}_\lambda$
satisfies (\ref{eq:unboundedness-along-imaginary}). Hence the function
$\mathcal{R}_\lambda$ and the sequences $\{\lambda_j\}_j$, $\{a_j\}_j$
satisfy the condition of Lemma~\ref{lem:auxiliary}. Thus, we have
shown the convergence of the series
\begin{equation}
  \label{eq:convergence-lemma}
 \sum_j\frac{\lambda_j^m}{\mathcal{R}_\lambda'(\lambda_j)}
\end{equation}
for all $m=0,1,2,\dots$, and that
\begin{equation}
  \label{eq:decomposition-lemma}
\frac{1}{\mathcal{R}_\lambda(\zeta)}=
\sum_j\frac{1}{\mathcal{R}_\lambda'(\lambda_j)(\zeta-\lambda_j)}\,.
\end{equation}
Taking into account Definitions 1 and 2 of the Addenda and Problems of
\cite[Chap.\,4]{MR0184042}, one obtains from Corollary 2 of
\cite[Addenda and Problems Chap.\,4]{MR0184042}, together with
conditions \emph{\ref{nec-suf-divergence}} and \emph{\ref{nec-suf3}},  that
$\{\lambda_k\}$ is a canonical sequence of nodes and
$\{a_k^{-1}\}$ the corresponding sequence of masses for the
moment problem given by $\{s_m\}_{m=0}^\infty$ with
\begin{equation}
  \label{eq:s-moments}
  s_m:=\frac{1}{\mathcal{M}}\sum_j\frac{\lambda_j^{m}}
  {\mathcal{R}_\mu(\lambda_j)\mathcal{R}_\lambda'(\lambda_j)}\,.
\end{equation}
Hence, for this moment problem, $\rho$ is a canonical solution
\cite[Def.\,3.4.1]{MR0184042}.  By definition, a canonical solution is
N-extremal and by \cite[Thm.\,2.3.3]{MR0184042}, the polynomials are
dense in $L_2(\mathbb{R},d\rho)$.

\item We prove that the elements of the basis
  $\{P_{n-1}(t)\}_{n=1}^\infty$ are in $D$. From
  (\ref{eq:convergence-lemma}) and (\ref{eq:decomposition-lemma}), by
  Lemma 1 of the Addenda and Problems of \cite[Chap.\,4]{MR0184042},
  one has for $m=0,1,2,\dots$
\begin{equation*}
   \sum_j\frac{\lambda_j^m}{\mathcal{R}_\lambda'(\lambda_j)}=0\,.
\end{equation*}
Then, if $S(t)$ is a polynomial
\begin{equation*}
  \langle(A_\rho+iI)S,\varphi\rangle_{L_2(\mathbb{R},d\rho)}
=\sum_j\frac{S(\lambda_j)}{\mathcal{R}_\lambda'(\lambda_j)}=0\,.
\end{equation*}
Whence it follows that $S\in D$.
\end{enumerate}
Now, by (\ref{eq:favard-system1}) and (\ref{eq:favard-system2}), it is
straightforward to show that $U^{-1}A_\rho\upharpoonright_DU$ (see
(\ref{eq:operator-u})) is a Jacobi operator in the limit circle case.

Denote by $J$ the Jacobi operator
$U^{-1}A_\rho\upharpoonright_DU$. On the basis of what was discussed
in Section~\ref{sec:preliminaries} one can find $\tau_1\in(\mathbb{R}
\cup\{\infty\})\setminus\{0\}$ such that the self-adjoint operator of
multiplication in $L_2(\mathbb{R},d\rho)$ is the canonical
representation of $J({\tau_1})$.  ${\tau_1}$ cannot be zero since then
$\{\lambda_k\}$ should contain the zero.
If $0\not\in\{\mu_k\}_k$, we define
\begin{equation}
  \label{eq:definition-of-g}
  {\tau_2}:=
  \begin{cases}
    \mathcal{M} & \text{if}\quad {\tau_1}=\infty\\
    \infty & \text{if}\quad {\tau_1}=-\mathcal{M}\\
    \frac{\mathcal{M}{\tau_1}}{{\tau_1}+\mathcal{M}} & \text{in all other cases},
  \end{cases}
\end{equation}
and if $0\in\{\mu_k\}_k$ simply assign ${\tau_2}:=0$.

For the proof to be complete it remains to show that $\{\mu_k\}_k$ are
the eigenvalues of $J({\tau_2})$. To this end we first show that
$\{\mu_k\}_k$ are the eigenvalues of some self-adjoint extension of
$J$. Let $\widetilde{\mathcal{M}}=-\mathcal{M}$ and define
\begin{equation*}
      \widetilde{a}_j:=\widetilde{\mathcal{M}}\mathcal{R}_\lambda(\mu_j)\mathcal{R}_\mu'(\mu_j)\,.
\end{equation*}
From condition \emph{\ref{nec-suf-moments}} with $m=0$, it follows that
$\widetilde{a}_j>0$ for any $j$ and $\sum_j\widetilde{a}_j^{-1}=1$,
and that the function $\widetilde{\rho}(t):=\sum_{\mu_k\le
  t}\widetilde{a}_k^{-1}$ is a solution of the moment problem
$\{s_k\}_{k=0}^\infty$ with $s_k$ given by
(\ref{eq:s-moments}). Moreover, taking into account conditions
\emph{\ref{nec-suf0}} and \emph{\ref{nec-suf3}}, one easily verifies
as before that the sequences $\{\mu_j\}_j$ and
$\{\widetilde{a}_j\}_j$, and the function $\mathcal{R}_\mu$
satisfy the conditions of Lemma~\ref{lem:auxiliary}. Therefore, by
Definitions 1, 2 and Corollary 2 of the Addenda and Problems of
\cite[Chap.\,4]{MR0184042}, as well as conditions
\emph{\ref{nec-suf-divergence}} and \emph{\ref{nec-suf3}},
it turns out that the sequence $\{\mu_j\}_j$ is a canonical
sequence of nodes and $\{\widetilde{a}_j\}_j$ the corresponding
sequence of masses for the moment problem given by
$\{s_k\}_{k=0}^\infty$ with $s_k$ satisfying (\ref{eq:s-moments}).
Hence $\widetilde{\rho}$ is a canonical solution of this moment
problem. Denote by $J(\widetilde{{\tau_2}})$ the self-adjoint extension of
$J$ having $\widetilde{\rho}$ as its spectral function.

Let us consider now the functions $R_{\tau_2}$ and $R_{\widetilde{{\tau_2}}}$
corresponding to $J({\tau_2})$ and $J(\widetilde{{\tau_2}})$, respectively (see
(\ref{eq:R-without-constants})).  It is straightforward to verify that
\begin{equation}
  \label{eq:g-g-tilde-eq}
  R_{\tau_2}(\lambda_j)=\frac{a_j}{\mathcal{M}R_{\tau_1}'(\lambda_j)}=
   R_{\widetilde{{\tau_2}}}(\lambda_j)\,,
\end{equation}
where the first equality follows from
(\ref{eq:normalizing-constants}), while the second follows from
(\ref{eq:R-derivative}) and (\ref{eq:sequence-def}).  By
(\ref{eq:constant-m}), (\ref{eq:calculation-M}), and
(\ref{eq:definition-of-g}), one easily concludes from
(\ref{eq:g-g-tilde-eq}) that ${\tau_2}=\widetilde{{\tau_2}}$.
\\
\end{proof}
\begin{remark}
  \label{rem:simlifications}
  When $0\in\{\mu_k\}_k$, the signs of the real numbers
  $\mathcal{R}_\mu(\lambda_j)\mathcal{R}_\lambda'(\lambda_j)$ and
  $\mathcal{R}_\lambda(\mu_j)\mathcal{R}_\mu'(\mu_j)$ are known. Thus,
  we can write condition \emph{\ref{nec-suf-constancy-sign}} as follows
  \begin{equation*}
    \mathcal{R}_\mu(\lambda_j)\mathcal{R}_\lambda'(\lambda_j)<0\,\quad
\mathcal{R}_\lambda(\mu_j)\mathcal{R}_\mu'(\mu_j)>0\quad\text{ for all
}j\,.
  \end{equation*}
  This is a consequence of (\ref{eq:C-0}) and (\ref{eq:constant-m}) by
  which we know that in equation (\ref{eq:normalizing-necessary})
  $M=-a(0)<0$.
\end{remark}
\begin{remark}
\label{rem:interlacing1}
  Note that, by Proposition~\ref{prop:interlacing}, conditions
  \emph{\ref{nec-suf0}}--\emph{\ref{nec-suf3}}\, imply the
  interlacing of the sequences $\{\lambda_k\}_k$ and $\{\mu_k\}_k$.
\end{remark}

\begin{acknowledgments}
  We thank A. Osipov for drawing our attention to \cite{MR997788} and
  the anonymous referees whose comments led to an improved presentation of our work.
\end{acknowledgments}

\end{document}